\newcommand{\anj}[2]{AJ, #1, #2}

\newcommand{\apjj}[2]{ApJ, #1, #2}
\newcommand{\aeta}[2]{A\&A, #1, #2}

\newcommand{\mn}[2]{MNRAS, #1, #2}

\newcommand{\nh}{N$_{\rm H}$}

\newcommand{\Halp}{H${\alpha}$}

\newcommand{\ergs}{erg s$^{-1}$}

\newcommand{\degree}{\degr}

\newcommand{\rxb}{\object{RX\,J0720.4-3125}}
\newcommand{\rxa}{\object{RX\,J1856.5-3754}}
\newcommand{\rx}{\object{RX\,J1605.3+3249}}
\newcommand{\prmo}{$\mu = 144.5\pm13.2$\,mas/yr}
\documentclass{aa}
\usepackage{graphicx}
\usepackage{epsfig}
\begin{document}
   \title{The proper motion of the isolated neutron star \rx  \thanks{Based on data collected at Subaru Telescope, which is operated by 
      the National Astronomical Observatory of Japan. } }

      \author{
       C. Motch 
      \inst{1}      
      \and
      K. Sekiguchi
      \inst{2}
      \and       
      F. Haberl
      \inst{3}
      \and
      V.E. Zavlin
      \inst{1,3}
      \and
      A. Schwope
      \inst{4}
      \and
      M.W. Pakull
      \inst{1}    
       }
   
   \offprints{C. Motch}

   \institute{
              Observatoire Astronomique, UMR 7550 CNRS, 11 rue de l'Universit\'e,
              F-67000 Strasbourg, France
	      \and
	      Subaru Telescope, National Astronomical 
	      Observatory of Japan, 650 North A'ohoku Place, Hilo, HI 96720, USA
              \and 
              Max-Planck-Institut f\"ur extraterrestrische Physik, D-85740,
              Garching bei M\"unchen, Germany 
	      \and
	      Astrophysikalisches Institut Potsdam, An der Sternwarte 16, D-14482 Potsdam, Germany
	      }
	      
   \date{Received ; accepted}

      \abstract{We obtained deep optical imaging of the thermally emitting X-ray bright and radio-quiet
isolated neutron star \rx \ with the Subaru telescope in 1999 and 2003. Together with  archival HST images
acquired in 2001 these data reveal a proper motion of \prmo . This implies a relatively high spatial
velocity and indicates that the star is unlikely to be re-heated by accretion of matter from the
interstellar medium. Assuming that \rx \ is a young (10$^{5}$-10$^{6}$\,yr) cooling neutron star, its
apparent trajectory is consistent with a birth in the nearby Sco OB2 OB association at a location close to
that derived for \rxa  \ and perhaps also to that of \rxb . This suggests that the X-ray bright part of
ROSAT-discovered isolated neutron stars is dominated by the production of the Sco OB2 complex which is the
closest OB association and a part of the Gould belt. The B and R magnitudes of the faint optical
counterpart did not vary from 1999 to 2003 at B = 27.22$\pm$0.10. Its B-R colour index of +0.32$\pm$0.17
is significantly redder than that of other isolated neutron stars and the optical flux lies a factor 11.5
above the extrapolation of the X-ray blackbody-like spectrum. The red optical colour reveals the presence
of an additional emitting component in the optical regime over the main neutron star thermal
emission. We also discovered a small elongated \Halp \ nebula approximately centered on the neutron star
and aligned with the direction of motion. The width of the nebula is unresolved and smaller than $\sim$
0.4\arcsec \ for a length of about 1\arcsec. The shape of the Balmer emitting nebula around \rx \ is very
different from those seen close to other neutron stars and should be confirmed by follow-up observations.
We shortly discuss the possible mechanisms which could give rise to such a geometry.}

      
   \titlerunning{The proper motion of RX\,J1605.3+3249}
   
   \maketitle
%

\section{Introduction}

A decade ago, it was expected that a large number of isolated neutron stars (INSs) accreting from
the interstellar medium would populate the ROSAT all-sky survey. It is now clear that this numerous
population is not detected, although a handful of X-ray bright INSs were found. These INSs are
radio-quiet, display thermal X-ray spectra with kT $\sim$ 40-100\,eV, undergo little interstellar
absorption indicating relatively small distances (few hundred parsecs) and are not associated with
any supernova remnant (see reviews in Treves et al. \cite{treves2000}, Motch \cite{mo2001} and
Haberl \cite{haberl03}). Four of them exhibit pulsations with periods around 10\,s (Haberl
\cite{haberl04}). Most of these INSs now have identified faint optical counterparts or good
candidates with B magnitudes in the range of 25.8 to 28.6. Their proximity and the apparent absence
of strong non-thermal activity makes them unique laboratories for testing radiative properties of
neutron star surfaces, high gravity and high magnetic field physics. This group of INSs could
eventually bring important constraints on the debated equation of state of matter in neutron
star interior. 

Grating X-ray spectra of the two brightest ROSAT-discovered INSs  \rxb\ (Paerels et al.
\cite{paerels2001}) and \rxa\ (Burwitz et al. \cite{bur2001}; Drake et al. \cite{drake02}) failed
to reveal any spectral feature that could be used to derive a gravitational redshift, for instance.
Their X-ray spectra are usually best fitted by a blackbody-like model. Further observations with
the EPIC and RGS instruments on board XMM-Newton have revealed in several cases the presence of
broad absorption lines with central energies in the 300-500 eV range on top of the blackbody
energy distribution. These structures have been tentatively interpreted as proton cyclotron lines
in magnetic fields of a few 10$^{13}$ G (Haberl et al. \cite{haberl03a}, Haberl et al.
\cite{haberl03b} and van Kerkwijk et al. \cite{vk2004}). A puzzling problem is that none of the models
fitting the X-ray data can adequately represent the optical-UV emission. Hydrogen atmosphere models
usually overpredict optical flux while the Rayleigh-Jeans tails of the X-ray blackbody model
typically fall short by a factor $\approx$ 10 at $\lambda$ $\approx$ 550\,nm (see Pavlov et al.
\cite{pavlov2002} for a recent review). Understanding neutron star surface emission properties
is therefore a prerequisite to derive reliable stellar radii. 

When observed, the X-ray pulsations can be interpreted as due to a non uniform temperature
distribution over the surface of a young cooling neutron star in strong magnetic field conditions or
to variable absorption lines.  The absence of detected pulsations in cases where evidence for strong
magnetic fields exists is also not well understood.

The evolutionary status of these objects, old accreting neutron stars or relatively young
cooling stars can be established by various means. A braking or an acceleration of the rotation period
can in principle reveal the powering mechanism (Zane et al. \cite{zane02}). This exercise is however
difficult and limited to pulsating stars. A potentially more efficient way is to search for proper
motion using either deep optical imaging or by taking advantage of the high spatial resolution
affordable by Chandra. Because of the strong dependency of Bondi-Hoyle accretion efficiency on
velocity relative to the interstellar medium, a high proper motion will naturally exclude the
accretion scenario, unless  high interstellar densities are locally encountered. Significant spatial
motions have been already detected from \rxa \ (Neuh\"auser~\cite{neu01}; Walter \cite{walter2001};
Kaplan et al. \cite{kaplan02b}) and from \rxb \ (Motch et al. \cite{motch03}) which in both cases
appear to exclude the accretion scenario. 

We report here on deep optical imaging obtained at the Subaru telescope and Hubble Space Telescope
of the optical counterpart of \rx \ (Motch et al. \cite{motch1999}) which is the third X-ray
brightest of its class. An optical counterpart has been identified by Kaplan et al.
(\cite{kaplan03a}). Our observations covering a four year interval of time reveal the high proper
motion of the object. A small but resolved \Halp \ nebula is also detected.

\section{Optical observations}

Optical imaging has been obtained on two occasions with the Subaru telescope in 1999 and 2003, and
in 2001 by the HST. 

Data from the first epoch have been retrieved from the SMOKA database at the end of January 2003.
These first observations were obtained between April 22 and May 15 1999 using the SuprimeCam
instrument (Miyazaki et al. \cite{miyazaki1998}) mounted at the Cassegrain focus. This
configuration which used a 3x2 CCD setup was in operation until July 1999 to perform an initial
detailed testing of both the telescope and instrument. Each ST-002A SITe CCD had 2048 x 4096 square
pixels with a size of 15$\mu$ corresponding to 0.03\arcsec \ on the sky at the Cassegrain focus.
For our purpose, we only used CCD \# 5 which fully contains the field of \rx . Of the 3 filters
used, B, R and broad \Halp \ , there exist only flat-fields for the R band. The \Halp\ filter
NA$\_$L651 has a central wavelength of 6521\AA\ and a full width of 325\AA . Due to relatively
large central wavelength variations with off axis angle (up to 100 \AA ), this filter was later
decommissioned. Laboratory measurements show that the \Halp\ line was in all cases well inside the
filter band pass but due to the spatial inhomogeneities the exact transmission at \Halp \ remains
uncertain. Individual exposure times were 1200\,s for the \Halp\ and B filters and 900\,s for the R
band. The telescope was moved by about 12\arcsec \ in both directions between two consecutive
exposures. All individual images were corrected for bias using over-scan areas and the R band
images were furthermore flat-fielded using well illuminated dome flats. Since only one bias
exposure exists for these observations, we prefered to correct images using mean over-scan values
in order to not increase photometric and positional errors. Individual images were then moved to
a common frame using a well exposed single reference star located close to the neutron star
position in order to minimize the effects due to the lack of correction for geometrical distortion.
The images were then stacked using a statistical cosmic-ray event rejection criterion and finally
rebinned with a 0.15\arcsec \ pixel size.

The HST archive provided second epoch observations. We used the unfiltered CCD (50CCD aperture)
image obtained on July 21 2001 which offers the highest sensitivity. Individual HST images were
geometrically corrected, drizzled onto a reference image with a pixel scale half of the original
one, filtered for cosmic-ray impacts and stacked together following the method outlined in
Fruchter \& Hook (\cite{fruchter2002}). These data were presented in Kaplan et al.
(\cite{kaplan03a}).

Third epoch images were all obtained on June 8 2003 with the FOCAS instrument (Kashikawa et al.
\cite{kashikawa2002}) operated at the Cassegrain focus. The target was located on CCD \# 2 which
has the best cosmetic quality. We obtained series of B and R band exposures with integration times
of 1200\,s and 600\,s respectively. Raw images were corrected for bias using average offset
exposures and flat-fielded with dome flats. A dedicated IDL procedure then corrected individual
images for geometrical distortion which otherwise could lead to errors of up to 2\arcsec \ at the
edge of the field of view. All images were moved to a common frame using a set of 9 reference
stellar-like objects and then stacked together using a statistical criterion for rejecting
cosmic-ray impacts. 

Ground based observations were in general obtained in photometric conditions although some thin
cirrus may have been present in 2003. The Subaru telescope is equipped with an atmospheric
refraction correction prism at the telescope side. The difference in colours between the relatively
red astrometric reference stars and the blue neutron star thus does not yield any significant
difference in position. 

The log of observations is listed in Table~\ref{obslog}. FWHM seeings are given as measured on the
summed images. 

\begin{table*}
\caption[]{Optical data}
\label{obslog}
\begin{tabular}{llcccccc}
\noalign{\smallskip}
Date           &  Instrument& Band    & Exp.  & Pixel         & FWHM image   & airmass\\
               &            &         & (ksec)&  (\arcsec )  & size (\arcsec )     & range\\
\hline
 22 April 1999 & Suprime-Cam &  \Halp  & 3.6      &  0.031\arcsec & 0.57\arcsec & 1.05 - 1.13\\  
 23 April 1999 & Suprime-Cam &  B      & 2.4      &  0.031\arcsec & 0.64\arcsec & 1.03 - 1.04\\    
 15 May  1999 & Suprime-Cam &  R      & 3.6      &  0.031\arcsec & 0.40\arcsec & 1.10 - 1.23\\
 21 July 2001 & HST/STIS    &  50CCD  & 2.7      &  0.0254\arcsec & 0.10\arcsec &            \\
  8 June 2003 & FOCAS       &  B      & 7.2      &  0.104\arcsec & 0.42\arcsec & 1.06 - 1.34\\
  8 June 2003 & FOCAS       &  R      & 4.8      &  0.104\arcsec & 0.58\arcsec & 1.50 - 2.40\\
\noalign{\smallskip}
\hline
\end{tabular}
\end{table*}

\section{Photometry}

\begin{figure}

\psfig{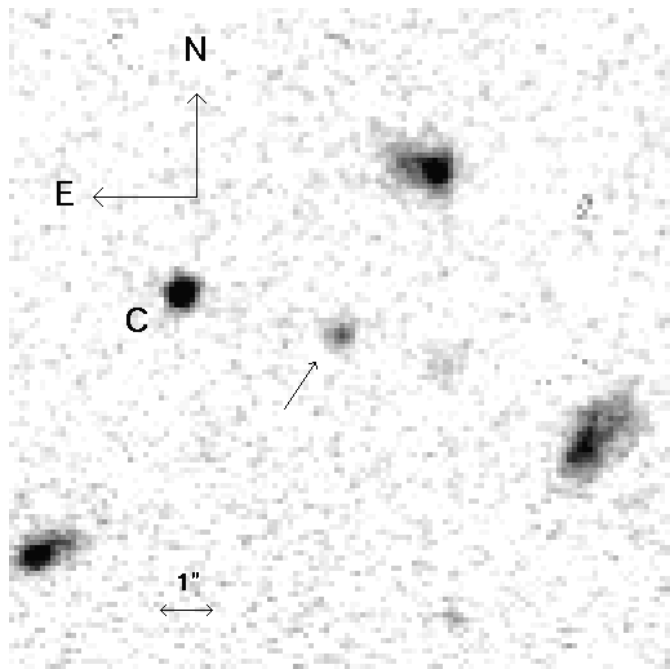}

\caption[]{ The summed B band image obtained in 2003 with FOCAS}

\label{sub03}
\end{figure}

During the second Subaru observations, B and R photometric standard stars were repeatedly observed
in the field of SA110-504 (Landolt \cite{landolt}). However, zero point variability by about 0.1
mag rms reveals the probable presence of thin cirrus during that night. We thus used our
photometric CFHT images obtained in 1998 and described in Motch et al. (\cite{motch1999}) to
calibrate several B and R local photometric reference stars in addition to those already mentioned
in that paper. Comparing CFHT and Subaru 2003 zero points shows that the 2003 images were indeed
absorbed by $\sim$ 0.15 mag in B and R. Magnitudes were computed using a 2-d Gaussian fitting
process taking into account possible sky background spatial variability. The magnitudes of \rx ,
listed in Table \ref{magrx} are generally consistent with those reported by Kaplan et al.
(\cite{kaplan03a}) from HST/STIS photometry. Errors quoted are one sigma and take into account
uncertainties in colour transformation and for the 1999 B value errors due to the absence of
flat-field correction. There is no evidence for flux variability on time scales of years. We show
in Fig. \ref{sub03} the summed B image obtained in 2003 at the Subaru telescope with FOCAS.

\begin{table}
\caption[]{B and R magnitudes of \rx}
\label{magrx}
\begin{tabular}{ccc}
\noalign{\smallskip}
Epoch           &   B               & R \\ 
\hline
April-May 1999  &  27.25 $\pm$ 0.19 & 26.90 $\pm$ 0.15 \\
June 2003       &  27.22 $\pm$ 0.10 & 26.90 $\pm$ 0.14 \\
\noalign{\smallskip}
\hline
\end{tabular}
\end{table}

Using the best determined value in 2003 we find B$-$R = +0.32 $\pm$ 0.17. This colour index is
significantly redder than that of \rxa \ (B$-$R = $-$0.61 $\pm$ 0.13, van Kerkwijk \& Kulkarni
\cite{vk2001}) and maybe also \rxb \ (B$-$R = $-$0.3 $\pm$ 0.4, Kulkarni \& van Kerkwijk
\cite{KvK98}). 

\section{Proper motion study}

Comparing positions obtained by three different instruments with different band passes requires
some caution. We estimated the errors resulting from uncorrected geometrical distortions by
measuring the dispersion in position of a number of bright stellar-like objects on individual
Subaru images. For the 1999 observations, the rms error is 12 mas in the B filter (4 objects)
and 7 mas in the \Halp \ and R filters (5 objects). Geometrical correction of the 2003 FOCAS
frames brings down the dispersion from about 12 mas to 4.6 mas. Therefore, the absence of available
distortion mapping for the Suprime-Cam instrument mounted at the Cassegrain focus has little
impact. However, the geometrical correction of FOCAS images indeed improved the astrometric
quality. 

We searched for proper motion using the sum of the B and R band images in 1999, the HST 50CCD image
and the B band FOCAS image in 2003 which provide the best compromise between sensitivity and image
quality. A group of astrometric reference objects were selected on the HST image for their
stellar-like profile or very small spatial extension. We used a 2-d Gaussian fitting process
allowing for sky gradient to measure accurate positions. Centering accuracies are listed in
Table \ref{cenac}. The Subaru 1999 and 2003 images were scaled and rotated to the HST frame using
9 and 7 such astrometric reference stars respectively. After transformation to the HST frame,
reference objects had mean position residuals of the order of 16 mas and 25 mas for the
Surprime-Cam and FOCAS, respectively. These residuals are larger than expected from centering
position errors only and probably reflect some remaining low level large scale field distortion.
The Subaru positional accuracies transformed to the HST reference frame take these systematic
errors into account.

\begin{figure}

\psfig{figure=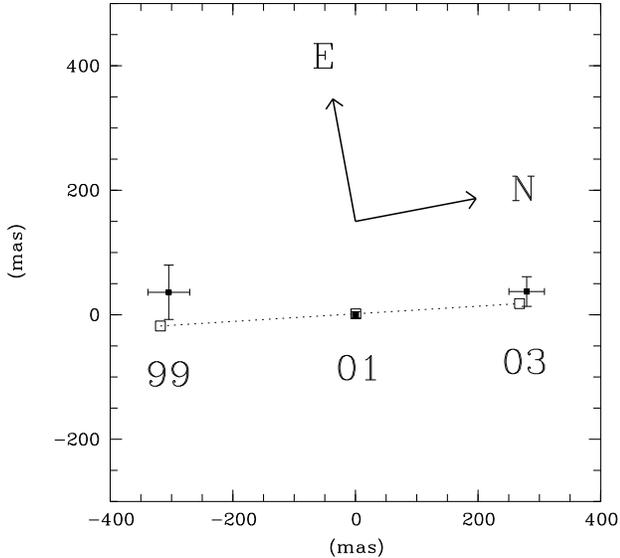,height=8cm,angle=-90,bbllx=10pt,bblly=10pt,bburx=590pt,bbury=620pt,width=8.8cm,clip=true}

\caption[]{ Positions of \rx \ for the 3 epochs (filled squares) plotted in the reference pixel
frame of the HST/STIS image. The best fit proper motion solution (\prmo ) is shown as open squares}

\label{pm}
\end{figure}

\begin{figure}

\psfig{figure=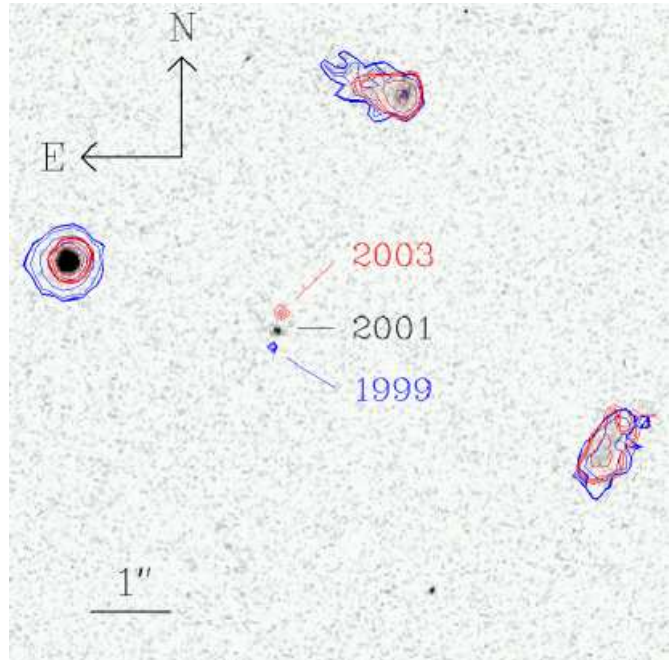,angle=0,width=8.8cm}

\caption[]{Contours of the Subaru observations plotted over the HST 50CCD image}.

\label{colouroverlay}
\end{figure}

A proper motion solution expressed in terms of total motion and direction was then fitted to the
three recorded positions of \rx \ using a minimum $\chi^{2}$ algorithm. The best fit shown in Fig.
\ref{pm} is obtained for \prmo \ with a direction of motion of  $-$9.86\degree \ $\pm$
5.65\degree\ eastward from north. This corresponds to
$\cos$($\delta$)$\mu_{\alpha}$=$-$24.7$\pm$16.3\,mas/yr and $\mu_{\delta}$=142.4$\pm$15.4\,mas/yr.
All errors are quoted here for a 90\% confidence level on one parameter. The detected proper motion
is thus highly significant and overall measurement and systematic errors appear negligible (see
Fig. \ref{colouroverlay}). The part of the apparent displacement due to solar motion should be
small, of the order of 5\,mas in right ascension and 11\,mas in declination at 100\,pc. 

Most of the proper motion of \rx \ is in galactic longitude ($\mu_{l}$ = 137 mas/yr). The source is
however slowly moving away from the galactic plane with $\mu_{b}$ = 37 mas/yr.  The unknown
distance and radial velocity of the neutron star does not allow us to derive the possible position
of its birth place with very good accuracy. In order to constrain the origin of \rx \ we assumed a
maximum age of $\sim$ 10$^{6}$\,yr, compatible with possible cooling times of a young neutron star
with kT $\sim$ 100\,eV (see below, Yakovlev \& Pethick \cite{yakovlev2004}), distances in the range
of 100\,pc to 500\,pc and radial velocities in the range of $\pm$ 700\,km\,s$^{-1}$. Considering
the short flight time, we also neglect the  effects of the galactic potential. The only nearby OB
association overlapping with some backward trajectories is the upper Scorpius-Centaurus region
belonging to the Sco OB2 complex. Amazingly, Walter \& Lattimer (\cite{wl2002}) derive a similar
birth place for \rxa \ in the upper Scorpius while Motch et al. (\cite{motch03}) speculate on a
possible birth place in the lower Centaurus-Crux part of the same Sco OB2 complex for \rxb .  

\begin{table}
\caption[]{Centering accuracies (mas) }
\label{cenac}
\begin{tabular}{lccc}
\noalign{\smallskip}
                &   Sub. 1999    &  HST 2001   &  Sub. 2003 \\    
\hline
Reference stars &   8 & 3 & 9 \\
\rx \           &  25 & 4 & 10 \\
\hline
\end{tabular}
\end{table}

\section{An \Halp \ nebula}

\begin{figure}

\psfig{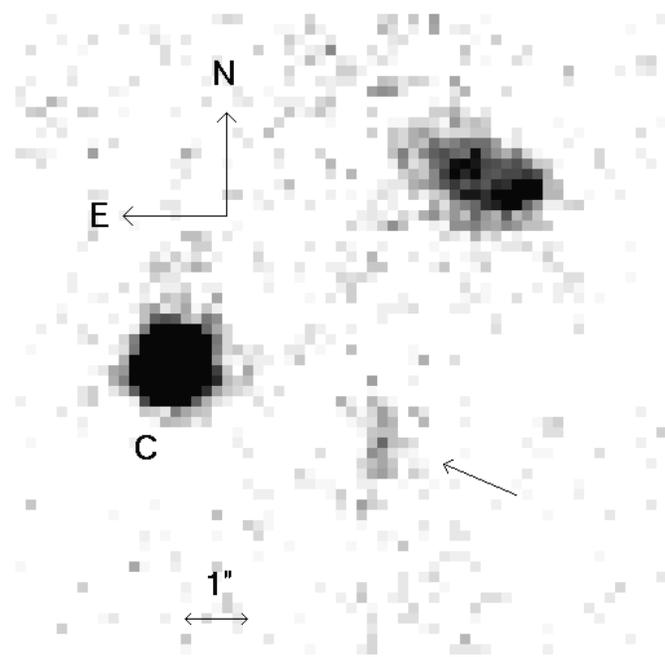}

\caption[]{ The \Halp \ image obtained in 1999 with the Suprime-Cam camera. The arrow
marks the neutron star position}

\label{halpha}
\end{figure}

We show in Fig. \ref{halpha} the \Halp \ image obtained in 1999. An extended and elongated source
is clearly seen at the position of \rx . The emission is detected on individual images whose
centers were shifted by several hundreds of pixels and is thus unlikely to be of instrumental
origin or caused by the absence of flat-field correction. The direction of largest extent of the
nebula is close to that of the apparent trajectory. \Halp \ nebulae have been discovered around a
small number of isolated neutron stars (see Chatterjee \& Cordes \cite{cc2002} for a recent
review). They have in general a well defined arc-like shape extending several tens of arcseconds on
the sky which can be interpreted as due to a bow shock created by the interaction of the high
velocity pulsar wind with the interstellar medium. In the case of \rxa \ \Halp \  emission could
also be due to  X-ray ionization (van Kerkwijk \& Kulkarni \cite{vk2001b}). The bright cylindrical
head of the Guitar nebula produced by the radio pulsar B2224+65 (Cordes et al. \cite{cordes93}),
although significantly larger, could be a scaled-up analogue of the nebula seen here. However, we
shall see below that this is in fact unlikely and that the shape and probably also the origin of
the nebula detected around \rx \ are unique. 

We tried to constrain the actual size and orientation of the \Halp \ emission by fitting to the
observed image a model consisting of a cylinder convolved with the point spread function. Owing to
the poor spatial resolution available, we assumed a simple uniformly emitting bar shape
characterized by its length, width and orientation. The Gaussian parameters of the point spread
function were derived from the nearby star "C" which is unresolved in the HST image and is a likely
dwarf M star (Motch et al. \cite{motch1999}). We used a 2-d minimum $\chi^{2}$ fitting process
taking into account CCD readout noise and counting statistics. The nebula seems slightly wider than
would be expected from the PSF alone. However, the resulting width of the bar depends on whether
the background is locally estimated or computed from a wider area as well as on the overall size of
the fitted image. The possible width ranges from 0\arcsec , i.e. unresolved, to $\approx$
0.4\arcsec . The best fit length is $\approx$ 1.1\arcsec \ with an inclination angle of 
$-$4.5\degree $\pm$ 5.5\degree \ (1$\sigma$ error) eastward from north, consistent with the
apparent direction of motion. 

\begin{figure}

\psfig{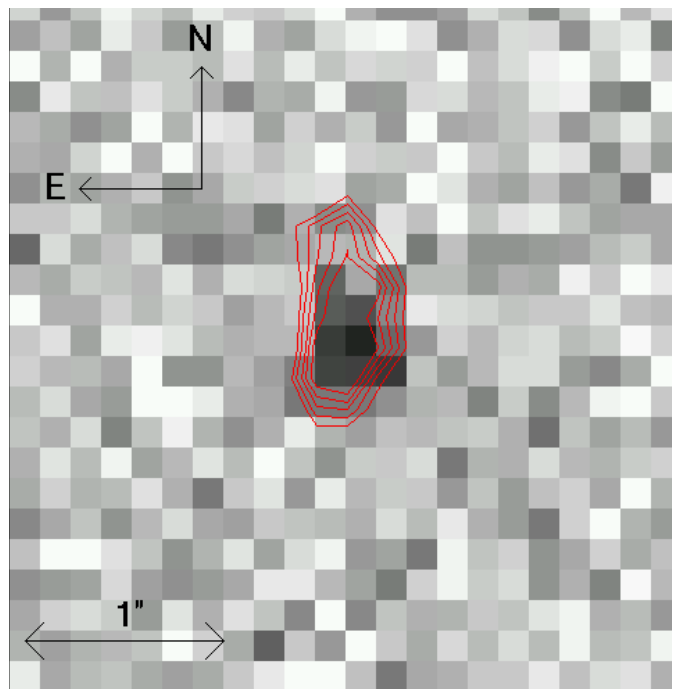}

\caption[]{ Overlay of the \Halp \ image on the R+B image 
obtained in 1999 with the Suprime-Cam camera. One pixel corresponds to 0.15\arcsec \ on the sky}

\label{haoverlay}
\end{figure}

The \Halp \ nebula is approximately centered on the neutron star (see Fig. \ref{haoverlay}) with a
slightly larger extent ($\approx$ 0.6\arcsec) toward the direction of motion. A total of 7 stellar
like objects were used to move the  \Halp \ image onto the combined R+B image with a rms positional
accuracy of 12 mas. The off center position of the neutron star with respect to its associated
nebula is thus significant. 

In the absence of flat-field and filter calibration at the position of \rx \ we can only
very roughly estimate the \Halp \ flux from the nebula. We assume a mean filter transmission of
80\% similar to that of the \Halp \ filter currently mounted on FOCAS and a total calibration error
of 20\%. The integrated flux from the nebula amounts to 2.4$\pm$0.6 10$^{-5}$ photon s$^{-1}$
cm$^{-2}$ or 7.2$\pm$1.8 10$^{-17}$ erg s$^{-1}$ cm$^{-2}$. Although much less extended than that
of \rxa \ (van Kerkwijk \& Kulkarni \cite{vk2001b}), the \Halp \ surface brightness of \rx \ seems
about 10 times brighter (for a 0.4\arcsec \ width) allowing its detection in a 1h exposure
only on an 8-m class telescope.

\section{Discussion}

\subsection{Proper motion}

\rx \ displays the second largest proper motion of the three objects of this class for which
measurements exist; \rxa \ has $\mu = 333\pm1$\,mas/yr (Neuh\"auser~\cite{neu01}; Walter
\cite{walter2001}; Kaplan et al. \cite{kaplan02b}) and \rxb \ has $\mu = 97 \pm$\,12 mas/yr (Motch
et al. \cite{motch03}). For \rx , a proper motion of \prmo \ corresponds to a transverse velocity
of V$_{\rm T}$ $\sim$ 68 $\times$ (d/100\,pc) km\,s$^{-1}$ with a small component due to solar
motion of $\sim$ 5.5$\times$(d/100\,pc)$^{-1}$\,km\,s$^{-1}$. 

A bolometric luminosity of $1.1\times 10^{31}$ (d/100\,pc)$^{2}$ \ergs\ (Motch et al.
\cite{motch1999})  derived from the blackbody fit to the ROSAT data, if powered by accretion from
the interstellar medium, would require a local particle density $\sim$ 170
(d/100\,pc)$^{5}$\,cm$^{-3}$. This furthermore assumes that there is no radial velocity component.
Such ISM densities are much larger than the average density toward the source; $n\, \sim \, $ 0.3
(d/100\,pc)$^{-1}$\,cm$^{-3}$, estimated from the hydrogen column density required by the blackbody
fit of the ROSAT X-ray spectrum (\nh $\sim$ 10$^{20}$ cm$^{-2}$, Motch et al. \cite{motch1999}). 
Following the arguments in Motch et al. (\cite{motch03}) we can conclude that like \rxb , \rx \ is
unlikely to be powered by accretion from the ISM unless extremely rare conditions are met. Similar
conclusions were reached by van Kerkwijk \& Kulkarni (\cite{vk2001b}) for \rxa \ based on the
modeling of its associated \Halp \ nebula.

\subsection{Is Sco OB2 the main provider of the brightest radio quiet INSs ?}

\rx \ is thus the third case of X-ray bright and radio quiet INS for which the accretion scenario
can be excluded with relatively high confidence. This confirms the conclusion that the thermally
emitting neutron stars detected by ROSAT are in general not powered by accretion from the
interstellar medium but are rather young cooling neutron stars with typical ages ranging from
10$^{5}$ to 10$^{6}$\,yr. The absence of accurate distance estimates, radial velocity information
and the still relatively large errors on proper motion direction of \rx \ and \rxb \ do not allow us
to draw unambiguous conclusions on the birth place of these neutron stars. There are however
relatively strong hints that the three X-ray brightest ROSAT discovered INSs, \rxa , \rx \ and
possibly \rxb \ were all born in the Sco OB2 association. With a mean distance of 120-140\,pc (de
Zeeuw et al. \cite{dezeeuw99}) this rich OB association is the closest to earth and belongs to the
large structure of young stars known as the Gould belt. Since most OB stars are found in
associations, we expect that statistically at least, the fact that a neutron star trajectory
crosses an OB association at a time roughly consistent with the cooling time is evidence for its
birth place. Some other middle-aged (age $\leq$ 4.25 Myr) nearby (d $\leq$ 350\,pc) neutron stars
have tentative birth places, Geminga possibly in Orion OB1 (Caraveo et al. \cite{caraveo1996}), PSR
J1932+1059 also in Sco OB2 (Hoogerwerf et al. \cite{hooger2001}). The revised distance of PSR
B0656+14 (d $\sim$ 290\,pc Brisken et al. \cite{brisken2003}) could also suggest a birth place in
Orion OB1. The tangential velocities of the X-ray bright and radio-quiet INSs have relatively
modest values for neutron star standards, 220$\pm$60 km\,s$^{-1}$ for \rxa \ (Kaplan et al.
\cite{kaplan02b}), 50 (d/100\,pc) km\,s$^{-1}$ for \rxb \ (Motch et al. \cite{motch03}) and 68
(d/100\,pc) km\,s$^{-1}$ for \rx . Their velocities are within the range observed for radio pulsars
and, assuming distances of $\sim$ 200\,pc for \rxb \ (Motch et al. \cite{motch03}) and $\sim$
300\,pc for \rx \ (Kaplan et al. \cite{kaplan03a}), suggest that they belong rather to the slow
part of the pulsar population (Arzoumanian et al. \cite{arzou2002}). 

Although still based on small number statistics, it seems that our current proximity to Sco OB2 may
be the reason why we can detect in relatively large numbers X-ray bright and radio quiet neutron
stars at small distances. In contrast, Orion OB1, the other major OB association in our
neighborhood may have provided Geminga and PSR B0656+14 but so far no ROSAT INS. Matching the
flight time from Orion (d $\sim$ 500\,pc) and the cooling time implies large radial velocities
($\sim$ 700 km\,s$^{-1}$). Such large radial velocities compared to tangential (120 km\,s$^{-1}$
for Geminga) could be understood as due to a selection effect since we only detect in X-rays the
neutron stars born in Orion which are young, hot and fast enough. The possibility is thus open that
radio quiet and hot neutron stars may preferentially be born with low kick velocities. However,
before such conclusions can be drawn, proper motions and distances of fainter objects must be
measured. 

\subsection{Optical to X-ray energy distribution}

\begin{figure}

\psfig{figure=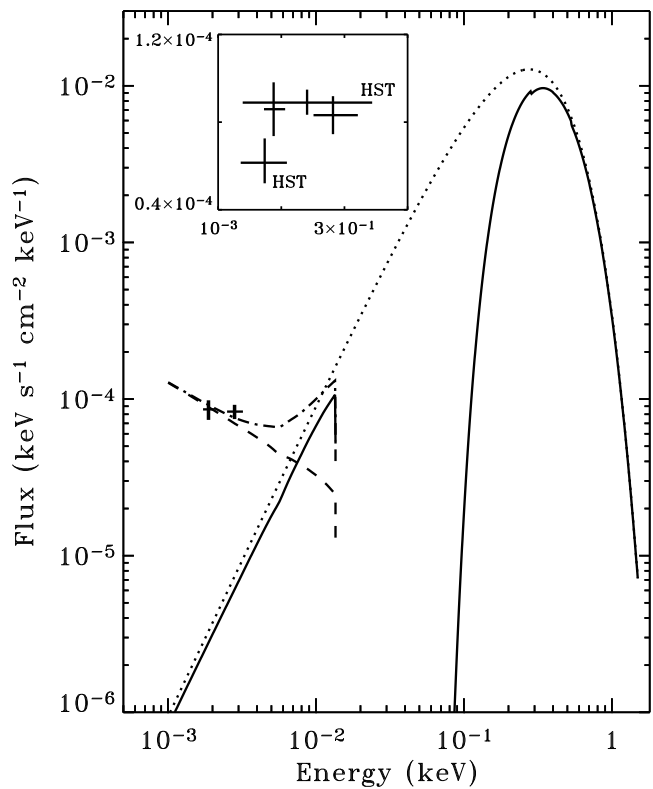,width=8.8cm,bbllx=135pt,bblly=300pt,bburx=457pt,bbury=705pt,clip=true}

\caption[]{The spectral energy distribution of \rx . Dotted line: unabsorbed blackbody fitting the
X-ray XMM-Newton EPIC pn data (see text). Thick line: corrected for absorption. The two data points
represent our B and R band measurements. The dashed curve is a power-law spectrum of photon index
1.53 and the dashed-dotted curve gives the sum of the blackbody and powerlaw spectra. Zoomed frame:
optical energy distribution based on HST and Subaru measurements}

\label{sed}
\end{figure}

We show in Fig. \ref{sed} the overall X-ray to optical distribution of \rx . The model describing
the X-ray continuum is a simple blackbody fit to EPIC pn data obtained in small window mode through
the thick filter during revolution 589. We preferentially used these data because of their low
pile-up fraction and high quality of the calibrations available for this instrumental setting
compared to others, especially the thin filter which suffers from some transmission uncertainties
at low energies. The blackbody fit gives a $\chi^{2}$ of 196 for 174 dof. Adding a Gaussian line
with a fixed energy of 0.45 keV  (van Kerkwijk et al. \cite{vk2004}) significantly improves the
fit, reducing the $\chi^{2}$ at 167.5 for 172 dof. The best fit Gaussian equivalent width is 29.5
eV with a width $\sigma$ $\sim$ 180\,eV. For simplicity we plot in Fig. \ref{sed} the best
blackbody-only fit with kT = 98.9\,eV and R$^{\infty}$ = 0.9 (d/100\,pc)\,km. 

As for all X-ray dim optically identified INSs, the optical continuum of \rx \ lies above the
Rayleigh-Jeans tail of the blackbody-like spectrum responsible for the X-ray energy distribution.
This has been sometimes used as an argument in favour of the existence of a second cooler thermal
component which would hardly contribute in X-rays but could account for most of the optical
emission (see Pavlov et al. \cite{pavlov2002}). In the case of \rxa \ , for which there exists an
accurate distance estimate based on a measured parallax, the low temperature implies a larger
radius of the emitting area more consistent with equations of state with conventional hadronic
matter (Burwitz et al. \cite{bur2003}). The two-blackbody model can naturally represent the strong
emissivity anisotropy induced by a large dipolar magnetic field. In this framework, the hot thermal
component would represent radiation emitted from the polar regions while the cooler one would be
emitted by the equatorial regions. However, in the case of \rx  \ (van Kerkwijk et al.
\cite{vk2004}) and even more acutely for \rxa \ (Ransom et al. \cite{ransom02}), the absence of
pulsations down to low pulse fractions puts severe constraints on the geometrical configuration and
amplitude of gravitational deflection. Standard hydrogen atmosphere models, while usually fitting
correctly the X-ray energy distribution, have the common problem to largely overpredict the optical
continuum. Thin hydrogen atmospheres such as proposed by Motch et al. (\cite{motch03}) for \rxb \
can account for the optical to X-ray energy distribution with a single component but have the
drawback of assuming the presence of an ad hoc amount of light elements on the surface of the
neutron star. A high surface reflectivity in the X-ray energy range such as in the case of
condensed atmospheres might also solve the X-ray - optical discrepancy (Burwitz et al.
\cite{bur2003}).

The B-R colour index of \rx \ (B$-$R = +0.32 $\pm$ 0.17) is significantly redder than that of \rxa
\ which essentially follows a F($\lambda$) $\sim$ $\lambda^{-4}$ continuum. Our B band flux is
consistent with that measured in the wide HST/STIS 50CCD band pass whereas our R band flux is
somewhat above that computed for the HST/STIS F28X50LP long pass filter. At present, the reasons
for this small discrepancy (formally a 2.1 $\sigma$ effect) are unclear. Due to flux calibration
uncertainties, the contribution of the nebula detected in the broad \Halp\ filter to the R band
flux is difficult to estimate. The fact that the nebula remains undetected in the R band images
suggests that the effect is not strong. The very small \nh\ required to fit the X-ray data (\nh\
$\leq$ 2.2 10$^{20}$ cm$^{-2}$, van Kerkwijk et al. \cite{vk2004}), corresponds to a maximum E(B-R)
of about 0.09, too small to account for the red colour index if an intrinsic Rayleigh-Jeans
spectrum is assumed.

The improved spectral resolution clearly shows that the energy distribution in the optical range
does not follow a Rayleigh-Jeans tail and reveals the presence of an additional component. A
similar situation exists in \rxb \ where a power law with an index slightly softer than
Rayleigh-Jeans is required to fit the optical/UV continuum (Kaplan et al. \cite{kaplan03b}) while
still keeping relatively blue overall colours. Generally, the B$-$R colour index of \rx \ appears
similar to that of two middle aged neutron stars, Geminga (B$-$R $\sim$ 0.2, Bignami et al.
\cite{bignami1993}) and  PSR B0656+14 (B$-$R = 0.62$\pm$0.11, Komarova et al. \cite{komarova2003}).
In these two pulsars, however, magnetospheric activity exists in the X-ray energy range (Zavlin
\& Pavlov \cite{zavlin04}). Mignami et al. (\cite{mignami1998}) find evidence for a peak at $\sim$
6000\AA \ which they interpret as the possible signature of cyclotron emission. For  PSR B0656+14,
a single power law component can in principle explain both the hard X-ray tail and the excess of
optical and near-infrared emission above a Rayleigh-Jeans component (Pavlov et al. \cite{pavlov2002}). 
Assuming that a similar mechanism is at work in \rx , we computed that a single powerlaw component
with a photon index of about 1.5, similar to that found in PSR B0656+14, could account for both the
2-10 keV EPIC pn flux 1 $\sigma$ upper limit (1.8 10$^{-14}$\,erg\,s$^{-1}$\,cm$^{-2}$) and for the
B to R band flux. The magnetospheric activity of \rx \ may thus well be a scaled down version of
that of PSR B0656+14.

\subsection{Nature of the \Halp \ nebula}

\Halp \ nebulae have been discovered around about six INSs, ordinary or millisecond pulsars. In all
cases an arc-shaped emission extending over several tens of arcseconds is visible with the main
nebula axis roughly aligned with the direction of proper motion (Chatterjee \& Cordes
\cite{cc2002}). Supersonic pulsar motion can create a bow shock nebula when the high velocity wind
which carries away most of the rotational kinetic energy interacts with interstellar medium. The
shock surface is located where ram pressure balance between the pulsar wind and the interstellar
medium takes place. \Halp \ emission then mainly arises from collisional excitation of neutrals by
shocked electrons and protons.  In many cases, the shape of the nebula can be reasonably well
fitted by the model of Wilkin (\cite{wilkin1996}) which assumes an isotropic wind and rapid
radiative cooling. Some asymmetric nebulae provide evidence toward density gradients in the ISM,
pulsar wind anisotropy or both (see e.g. Gaensler et al. \cite{gaensler2002}). 

No such arc-shaped nebula is seen around \rx . Our relatively low detection sensitivity could
explain part of the observed differences. However, in the known bow shock nebulae Balmer emission
usually shows the highest surface brightness ahead of motion. At the apex the nebula extends
perpendicular to the trajectory thus tracing the shape of the front-shocked region. In the present
case, the \Halp \ emission appears elongated in the direction of motion. Such a geometry cannot be
matched by any isotropic wind model since one would naively expect a larger extent in the direction
perpendicular to the motion than in the direction of motion. In the absence of detailed modeling,
it is unclear whether pulsar wind beaming could account for the observed geometry.

If no strong wind blows from \rx \ \Halp \ emission could still arise from an X-ray ionized nebula.
Blaes et al. (\cite{blaes1995}) have shown that ionization of the local interstellar medium by the
X-ray and UV radiation field of the hot neutron star combined with proper motion would produce
cometary HII regions. Their properties depend on the details of the ionization, recombination,
heating and cooling rates (see van Kerkwijk \& Kulkarni \cite{vk2001b}, for a detailed modeling
including dynamical effects). The isotropic ionizing field should create a spherical head nebula
centered on the neutron star with an extent in the direction perpendicular to motion at least as
large as in the direction of motion, similar to the case of bow shock nebulae. Also, because of the
rather long cooling time, most of the \Halp \ emission will occur in the wake of the neutron star,
again at variance with the observation. Scaling the observed \Halp \ surface brightness to that of
\rxa \ (van Kerkwijk \& Kulkarni \cite{vk2001b}) would imply local densities of the order of
$n_{\rm H} \ \sim$ 9 (d/100\,pc)$^{-1}$\,cm$^{-3}$, much larger than expected for normal ISM
densities.  Such high densities would however yield a cooling time of $\tau_{cool}$ $\sim$
6\,(d/100\,pc)\, yr roughly comparable to the nebula crossing time $\tau_{cross}$ $\sim$ 4\,yr as
estimated from its angular size ($\sim$ 600\,mas ahead) and observed proper motion. 

The geometry of the nebula is in principle consistent with that expected from a bipolar jet. Such
jets have been observed in X-rays from very young neutron stars, the best documented cases being
those of the Crab (Weisskopf et al. \cite{weisskopf2000}) and of the Vela pulsar (Helfand et al.
\cite{helfand2001}). Their power-law X-ray spectra are interpreted as due to optically thin
synchrotron emission from ultra-relativistic particles. Obviously, some other mechanism must be
invoked to explain \Halp \ emission at redshift $\leq$ 0.03 (the maximum allowed by the filter band
pass) from such a jet. One possibility could be the interaction of the beamed particle flow with
the interstellar medium. If the main direction of the \Halp \ nebula would indeed trace the spin
axis, \rx \ would be another case of apparent alignment of spin axis and proper motion direction
(Lai et al. \cite{lai2001}).

Alternatively, if the main axis of the nebula would exactly follow the trajectory, as would be the
case for an X-ray ionized or bow shocked nebula unperturbed by the interstellar medium, the
apparent proper motion direction should not be exactly aligned with the nebula as a result of solar
motion. In principle, this effect could be used to derive independent estimates of the distance of
the neutron star. However, the accuracy with which we can determine the angle difference $\theta$ =
4.6\degree $\pm$ 6.3\degree \ is not yet large enough to be useful. Nevertheless, the sign of the
angle difference between the nebula and the apparent proper motion is that expected due to the
different positions of the Sun between 1999 and 2003. Using the Sun velocity parameters determined
by Hipparcos (Dehnen \& Binney \cite{db98}) we compute an angle difference $\theta$ = 1.3\degree
(d/100\,pc)$^{-1}$ .

Considering the absence of calibration of the instrumental \Halp \ setting, it is clear that the
astonishing geometry of the nebula needs to be confirmed and specified by other observations before
extensive conclusions on its nature can be drawn.

\section{Conclusions}

Our deep Subaru imaging reveals the high proper motion of \rx \ and thus confirms the optical
identification proposed by Kaplan et al. (\cite{kaplan03a}). The corresponding large space motion
most probably excludes accretion from the interstellar medium as the source of the radiated X-ray
energy. Similarly to \rxa \ and \rxb, \rx \ is likely a cooling neutron star with an age younger
than $\sim$ 10$^{6}$ yr. Evidence for residual magnetospheric activity comes from the relatively
red B$-$R index which indicates the presence of an excess of red light above the expected
Rayleigh-Jeans tail, a picture comparable to that seen in Geminga or in PSR B0656+14. A small size
\Halp \ nebula is detected approximately centered on the neutron star and extended along the
direction of the trajectory. The lack of photometric calibration and flat-fielding does not make it
possible to draw stringent conclusions on its nature. However, the small size and geometry of the
nebula are unique and no standard bow-shock or X-ray ionisation model can account for these
peculiarities in a straightforward manner. Forthcoming deeper observations should allow us to
confirm the unusual features detected in 1999. Finally, we find that \rx \ may have been born in the same
nearby Sco OB2 OB association as \rxa \ and also perhaps \rxb . This could imply that the bright
part of the observed LogN-LogS curve for X-ray dim isolated neutron stars is dominated by the
production of this single OB association.

\begin{acknowledgements}

We thank the Subaru telescope team for providing access to the instrument and for help during the
observations. We are particularly indebted to Dr. Yutaka Komiyama and Dr. Masafumi Yagi for
carrying out the 1999 observations. Part of the data presented here were based on observations
made with the NASA/ESA {\it Hubble Space Telescope}, obtained at the Space Telescope Institute,
which is operated by the Association of Universities for Research in Astronomy, Inc. under NASA
contract NAS 5-26555. We thank an anonymous referee for useful comments.

\end{acknowledgements}

\end{document}